# Drastic suppression of CDW (charge density wave) by Pd addition in $TiSe_2$

Abhilasha Saini[1,2], N. K. Karn[1,2], Kapil Kumar[1,2], R.P. Aloysius[1,2] and V.P.S. Awana[1,2,*]

[1]*CSIR-National Physical Laboratory, Dr. K. S. Krishnan Marg, New Delhi-110012, India*
[2]*Academy of Scientific and Innovative Research (AcSIR), Ghaziabad 201002, India*

**Abstract:**

$TiSe_2$ is a known Topological semimetal (TSM) having both the semi-metallic and topological characters simultaneously along with the Charge density wave (CDW) at below 200K. In the current short article, we study the impact of Pd addition on CDW character of $TiSe_2$ and the possible induction of superconductivity at low temperatures. Bulk samples of $TiSe_2$ and $Pd_{0.1}TiSe_2$ are synthesized by solid-state reaction route, which are further characterized by powder X-ray diffraction (PXRD) and field emission scanning electron microscopy (FESEM), respectively, for their structural and micro-structural details. The vibrational modes of both samples are being analyzed by Raman spectroscopy, showing the occurrence of both $A_{1g}$ and $E_g$ modes. CDW of pure $TiSe_2$ seen at around 200K in electrical transport measurements, in terms of sharp semi-metallic to metallic transition peak with hysteresis in cooling/warming cycles is not seen in $Pd_{0.1}TiSe_2$ and rather a near metallic transport is seen down to 2K. Although superconductivity is not seen down to 2K, the CDW transition is seemingly completely suppressed in $Pd_{0.1}TiSe_2$. It is clear that Pd addition in $TiSe_2$ suppresses CDW drastically. Trials are underway to induce superconductivity in Pd-added $TiSe_2$. Density functional theory (DFT) calculations show primary evidence of suppression of CDW by addition of Pd in $TiSe_2$ due to an increase in the density of states.

[*]**Corresponding Author**
Dr. V. P. S. Awana: E-mail: awana@nplindia.org Ph. +91-11-45609357, Fax-+91-11-45609310
Homepage: awanavps.webs.com

**Introduction:**

The charge density wave (CDW) is an exotic phase of the quantum matter and is considered to be a symmetry-broken state of the same. This interesting electronic phase contains the spatially periodic variation in the density of conduction electrons, which frequently results in a dramatic renormalization of the electronic spectrum, including the appearance of energy gaps [1-3]. $TiSe_2$ is one of the extensively studied CDW materials, which is a member of the family of Group IV transition metal dichalcogenides (TMDCs). TMDCs contain covalently bonded trilayers in which, transition metal layer is in the middle of two chalcogen layers [4-6]. These trilayers are stacked and hold together by weak Vander Waals forces [7-9].

Doped TiSe$_2$ shows variety of properties as the same shows Kondo effect on doping of 3d transition elements due the presence of spin flip centers in the TiSe$_2$ unit cell [10]. Electronic state of TiSe$_2$ is debatable as it is reported to have one of two typical states: either a semimetal [11] or a semiconductor [12] with a modest indirect gap. TiSe$_2$ shows CDW transition below 200 K, with no intermediate incommensurate phase [13-14]. More interestingly, TiSe$_2$ is found to show superconductivity with doping of Cu or Pd atoms, which is accompanied by suppression of CDW [15,16]. Intercalation and chemical substitution are the most suited routes to alter the electron counts, which further adjust the CDW transition in TiSe$_2$ [15,16]. Here, we present the drastic suppression of CDW in 1T-TiSe$_2$ caused by doping with electrons through intercalation of Pd in TiSe$_2$. Density functional theory (DFT) is very promising tool in predicting and confirming the materials properties by computational simulations [17]. We have calculated the band structure and density of states for pure and Pd added TiSe$_2$. In this short letter, we report synthesis of polycrystalline Pd$_{0.1}$TiSe$_2$. The synthesized sample is examined in context of phase purity through and elemental composition through XRD and energy dispersive X-ray analysis (EDAX) techniques, respectively. Raman modes are also observed for TiSe$_2$ and Pd$_{0.1}$TiSe$_2$ through Raman spectroscopy. The signature feature of TiSe$_2$ i.e. CDW, is found to be marginal or nearly non-existent in Pd$_{0.1}$TiSe$_2$. The characteristic hysteresis in the cooling and warming cycle observed earlier [18] in ρ-T measurements is suppressed in Pd$_{0.1}$TiSe$_2$. Further, we present primary evidence of suppression of CDW by addition of Pd in TiSe$_2$ due to an increase in the density of states.

**Experimental:**

TiSe$_2$ and Pd$_{0.1}$TiSe$_2$ polycrystalline samples were synthesized utilizing a simple self-flux technique as reported in ref. 18. Ti (99.99%), Se (99.99%), and Pd (99.99%) powders were taken in stoichiometric proportions and were completely ground in an argon-filled MBRAUN glove box. The acquired mixed powders were compressed into a rectangular pellet using a hydraulic press, and it was then vacuum encapsulated in a quartz tube at around $5\times10^{-5}$ Torr vacuum. The vacuum-encapsulated samples were put into PID controlled muffle furnace and were heat treated as per the detailed protocol given in ref. 18. For clarity, the schematics for both stages of the polycrystalline TiSe$_2$ and Pd$_{0.1}$TiSe$_2$ synthesis process are shown in Figures 1(a) and (b).

Rigaku Miniflex II X-ray diffractometer fitted with Cu-Kα radiation of wavelength 1.5418Å is used to record XRD pattern for the structural analysis and phase purity determination. Rietveld refinement of PXRD patterns is performed using Fullprof software, and unit cells are extracted using VESTA software, which utilizes refined parameters from Rietveld analysis. Zeiss EVO-50 FESEM is used to determine surface morphology and EDAX measurement. Raman spectra of bulk TiSe$_2$ and Pd$_{0.1}$TiSe$_2$ are taken at room temperature using the Jobin Yuvon Horiba T64000 Raman Spectrometer for a wave number range of 100-350 cm$^{-1}$. The magneto-transport measurements of the produced TiSe$_2$ and Pd$_{0.1}$TiSe$_2$ polycrystalline sample were performed utilizing the Quantum Design Physical Property Measurement System (QD-PPMS), with a conventional four-probe approach.

**Results & Discussion:**

The PXRD pattern of synthesized $TiSe_2$ and $Pd_{0.1}TiSe_2$ polycrystalline samples are shown in Fig. 2(a). The PXRD patterns of synthesized $TiSe_2$ and $Pd_{0.1}TiSe_2$ polycrystalline samples are well-fitted with the parameters of a trigonal crystal structure with P-3m1(164) space group. All of the peaks can be indexed with their respective planes in the applied unit cell parameters, suggesting that the synthesized sample is phase pure and visibly no impurity is present in the samples. The parameter of the goodness of fit, i.e., $\chi^2$ is found to be 5.63 and 4.34 for $TiSe_2$ and $Pd_{0.1}TiSe_2$, these values are in the acceptable range indicating the quality of Rietveld refinement of PXRD data. The Rietveld refined unit cell parameters are a=b=3.549(2)A, c=6.011(4)A and a=b=3.581(2)A, c=5.972(4)A respectively for $TiSe_2$ and $Pd_{0.1}TiSe_2$. The lattice parameters of $Pd_{0.1}TiSe_2$ differs from pure $TiSe_2$ as both the a and b parameters are increased with a slight decrement in c parameter. This shows that the Pd intercalation in $TiSe_2$ unit cell results in contraction along c axis, with a slight expansion along ab plane. These results are in agreement with ref. 16 on Pd added $TiSe_2$. Crystallographic information files (CIF) obtained from Rietveld refinement is processed in VESTA software to draw unit cell of synthesized samples and the same are shown in fig. 2(b) and (c). It is obvious from unit cells that the investigated $TiSe_2$ and $Pd_{0.1}TiSe_2$ polycrystalline sample displays a layered structure separated by Vander Waals gap. The presence of Vander Waals gap provides ample opportunity to intercalate foreign atoms in $TiSe_2$ unit cell. In Fig. 2(c) Pd atoms in $Pd_{0.1}TiSe_2$ are shown to reside in Vander Waals gap of $TiSe_2$ as Pd does not substitute Ti in the same.

Fig. 3(a) exhibits FESEM image of synthesized $Pd_{0.1}TiSe_2$ polycrystalline sample. FESEM image shows the presence of grains, which are separated by grain boundaries confirming polycrystalline nature of synthesized sample. Fig. 3(b) shows the bar chart of the elemental composition of the synthesized polycrystalline sample $Pd_{0.1}TiSe_2$. EDAX measurements confirm the presence of Pd atoms, which shows that the same is successfully intercalated in $TiSe_2$ unit cell. All the constituent elements viz. Pd, Ti and Se are found be in close proximity to stoichiometric ratios. The FESEM details of pristine $TiSe_2$ are reported earlier in ref.18.

Raman spectra of $TiSe_2$ and $Pd_{0.1}TiSe_2$ polycrystalline samples are recorded at room temperature and the same are shown in Fig. 4. Raman modes are observed at 150 cm$^{-1}$, 255 cm$^{-1}$ and 155 cm$^{-1}$, 250 cm$^{-1}$ respectively for pure $TiSe_2$ and $Pd_{0.1}TiSe_2$. These vibrational modes are identified as the $A_{1g}$ and $E_g$ modes, and are in agreement with the prior study on the 1T phase of $TiSe_2$ [19]. The schematic of the observed vibrational modes is given in the inset of Fig. 4. $E_g$ mode indicates a doubly degenerate phonon mode in which the Se atoms move along the in- plane directions, $A_{1g}$ mode represents the phonon mode in which two Se atoms move along the out-of-plane directions of $TiSe_2$. Raman modes are observed to be slightly shifted for $Pd_{0.1}TiSe_2$ as compared to pure $TiSe_2$ due to intercalation of Pd atoms, which results in change in unit cell parameters. The shift in Raman modes also confirms the intercalation of Pd atoms in TiSe$_2$ unit cell in Pd$_{0.1}$TiSe$_2$.

Fig. 5(a), (b) and (c) show the resistivity versus temperature (ρ-T) measurement results of synthesized TiSe$_2$ and Pd$_{0.1}$TiSe$_2$ polycrystalline samples. TiSe$_2$ is found to show an intriguing anomaly in temperature range 150K to 200K due to ordering of charge carries. This charge carrier ordering occurs due to the presence of CDW phase in pure TiSe$_2$, and is a well-known phenomenon of the same [20-23]. Both warming and cooling data show a hysteresis in CDW, showing the same to be a first order phase transition. TiSe$_2$ exhibits dual electronic phase as seen in Figure 5 (a), as the temperature is lowered; the insulating phase begins to take precedence over the metallic phase. The metallic phase remains present despite the insulting phase being highest at CDW, and this presence of two electronic phases causes hysteresis in the warming and cooling cycle. Fig. 5(b) shows ρ-T measurement results of synthesized Pd$_{0.1}$TiSe$_2$ polycrystalline samples. Pd doping in TiSe$_2$ results in dramatic suppression in CDW transition, which is evident from Fig. 5(b). The ρ-T plot of Pd$_{0.1}$TiSe$_2$ is found to be similar to normal metals. A small bump has been observed in ρ-T plot in temperature range 200K-250K, which can be regarded as a suppressed CDW transition; this is in agreement with the previous report [16]. The enhanced charge carrier concentration due to intercalation of Pd atoms can be the reason of suppression of CDW phase in the same. The CDW transition occurs at higher temperature as compared to pure TiSe$_2$. A small hysteresis is also observed around the resistivity anomaly signifying the presence of a first order phase transition. The smaller hysteresis in resistivity in cooling and warming cycle in Pd$_{0.1}$TiSe$_2$ signifies the much reduced dual electronic conductivity nature in the same as compared to pure TiSe$_2$, which is evident from nearly metallic ρ-T plot. Interestingly, Pd$_{0.1}$TiSe$_2$ is reported to show superconductivity below 2K [16], found to be absent here as no sign of resistivity drop is observed down to 2K (lowest temperature limit of our PPMS), though the possibility of occurrence of superconductivity at temperature below 2K cannot be neglected here. Further, low temperature measurements below 2K are required to get more insight of superconducting phase in Pd$_{0.1}$TiSe$_2$. Fig. 5(c) shows the comparison of normalized ρ-T plot of both pure TiSe$_2$ and Pd$_{0.1}$TiSe$_2$. It is clear from Fig. 5(c), that the CDW phase in Pd$_{0.1}$TiSe$_2$ nearly disappeared in comparison to pure TiSe$_2$ and the same seems to be in normal metallic phase. In pure TiSe$_2$, the resistivity value is found to increase 3 times of room temperature resistivity near CDW transition (170K), while the same is not visible in Pd$_{0.1}$TiSe$_2$.

Previously there are several rigorous first-principle studies on TiSe$_2$ [24-26]. To understand the suppression in CDW due to Pd addition in TiSe$_2$, we compute the electronic band structure and Density of states for pure TiSe$_2$ and Pd intercalated TiSe$_2$ within the DFT framework. From the previous report [15], Pd atom is intercalated in between the 2D layers of TiSe$_2$. To simulate the intercalation, a supercell of size 3×3×1 created, and a Pd atom is inserted in between the TiSe layers, as shown in Fig. 6. Thus for simulation, we have Pd ~11% in TiSe$_2$. All computational simulation was performed in Quantum Espresso software based on density functional theory (DFT) [27, 28]. First, self-consistent calculations were done using the experimental lattice parameters for pure and Pd intercalated samples. After that, the bulk electronic band structure and Density of states were calculated. For DFT calculation, GGA incorporated Perdew-Burke-Ernzerhof (PBE) type ultrasoft pseudopotentials are used to include the effect of electronic exchange and correlations corrections. For band structure calculation first Brillouin Zone (BZ) is discretized on a mesh of 11×11×5 for pure and 4×4×6 for doped systems given by Monkrost-Pack. For the convergence of self-consistent calculation, the cutoff is 1.2×10-9 Ry, and the charge cut-off of 400 Ry and wave function cut-off of 48 Ry is used. The total energy convergence criteria is set to be $4\times10^{-4}$ eV/atom.

For the input of band structure calculations, unit cell parameters were taken from Reitveld refined structures. The k-path followed for band calculations is determined from the SeeK-path: the k-pathfinder and visualizer [29]. The suggested optimized path is Γ →M →K →Γ →A →L →H →A. Figs. 7(a) and 7(b) show the electronic band structure for the pure and Pd intercalated $TiSe_2$ respectively. Figs. 7(c) and 7(d) show the DoS for the pure and Pd intercalated $TiSe_2$ respectively. We observe a substantial change in the band structure as well as in DoS near the Fermi level. In Pd intercalated $TiSe_2$, the near Fermi level bands are dense, indicating higher DoS. Same results we find by directly comparing the DoS of pure and Pd intercalated $TiSe_2$. Increased DoS near the fermi-level implies the higher number of charge carriers, which is possibly the reason of observed decrease in CDW. The substantial change in DoS also confirms the semi-metal-to-metal transition due to Pd addition.

**Conclusion:**

In this study, we synthesized polycrystalline samples of $TiSe_2$ and $Pd_{0.1}TiSe_2$, which are well characterized using PXRD, FESEM, and Raman spectroscopy. Intercalation of Pd atoms in $TiSe_2$ unit cell is confirmed by change is lattice parameters and shift in Raman modes. Pd atoms are found to reside in the voids of $TiSe_2$ unit cell. Intercalation of Pd atoms is shown to have a large impact on transport properties, as the CDW transition is completely suppressed in comparison to pure $TiSe_2$. We observed no superconducting transition down to 2K. The theoretical simulations show that suppression in CDW is directly related to the increment in DoS due to Pd addition in $TiSe_2$.

**Acknowledgment:**

The authors would like to thank the Director of National Physical Laboratory (NPL), India, for providing the facilities and his keen interest in research. Authors would like to thank Dr. J.S. Tawale and Ms. Shaweta for FESEM and Raman spectroscopy measurements respectively and Mr. Manish Mani Sharma for help in resistivity measurements and critical reading of the short MS. Abhilasha Saini, Kapil Kumar and N. K. Karn would like to thank AcSIR-NPL for Ph.D. registration.

**Figure Captions:**

**Fig. 1(a)** Schematic of 1$^{st}$ step of heat treatment followed to synthesize $TiSe_2$ and $Pd_{0.1}TiSe_2$ polycrystalline sample. **(b)** Schematic of 2$^{nd}$ step of followed heat treatment.

**Fig. 2(a):** Rietveld refined PXRD of polycrystalline $TiSe_2$ sample and $Pd_{0.1}TiSe_2$ **& (b)** Unit Cell Structure through VESTA software of $TiSe_2$ and $Pd_{0.1}TiSe_2$.

**Fig. 3:** FESEM image of synthesized $Pd_{0.1}TiSe_2$ polycrystalline sample and (b) the atomic composition of constituent elements.

**Fig.4:** Room Temperature Raman Spectra of $TiSe_2$ and $Pd_{0.1}TiSe_2$ polycrystalline in which inset is showing the schematic of $TiSe_2$ and $Pd_{0.1}TiSe_2$ Raman Modes ($A_{1g}$ & $E_g$).

**Fig. 5(a):** $\rho$-T measurements of $TiSe_2$ polycrystalline sample in warming and cooling cycles **(b)** $\rho$-T measurements of $Pd_{0.1}TiSe_2$ polycrystalline sample in warming and cooling cycles **(c)** Normalized $\rho$-T measurements of both $TiSe_2$ and $Pd_{0.1}TiSe_2$.

**Fig. 6:** Pd intercalated $TiSe_2$ constructed on a supercell of 3×3×1.

**Fig. 7:** DFT calculated Electronic band structure of **(a)** pure $TiSe_2$ **(b)** Pd intercalated $TiSe_2$. DoS of **(c)** pure $TiSe_2$ **(d)** Pd intercalated $TiSe_2$ showing a substantial increase in DoS due to Pd intercalation in $TiSe_2$.

**References:**

1. Rossnagel K., J. Phys., Condens. Matter, **23**, 213001 (2011).
2. Hellmann S., Rohwer T., Kallane M., Hanff K., Sohrt C., Stange A., Carr A., Murnane M. M., Kapteyn H. C., Kipp L., Bauer M. and Rossnagel K., Nat. Commun., **3**, 1069 (2012).
3. Q. Hu , J. Y. Liu, Q. Shi, F. J. Zhang, Y. Zhong, L. Lei and R. Ang, EPL, **135**, 57003 (2021).
4. Alexander A. Balandin, Sergei V. Zaitsev-Zotov and George Gruner, Appl. Phys. Lett. **119**,170401 (2021).
5. Davied B. Lioi, David J. Gosztola, Garry P. Wiederrecht and Goran Karapetrov, Appl. Phys. Lett. **110**,081901 (2017).
6. Sajedeh Manzeli, Dmitry Ovchinnikov, Diego Pasquier, Oleg V. Yazyev and Andras Kis, Nat Rev Mater **2**, 17033 (2017).
7. A.N. Titov, Yu. M. Yarmoshenko, P. Bazylewski, M.V. Yablonskikh et al. Chemical Physics Letters **497,** 187 (2010).
8. M.B. Dines., Science **188**, 1210 (1975).
9. C. H. Lee et al., Nat. Nanotechnol. **9**, 676 (2014).
10. M. Sasaki,A.Ohnishi,T. Kikuchi, M. Kitaura,Ki- Seok Kim and Heon-Jung Kim, Phys. Rev. B **82**, 224416 (2010).
11. H. N. S. Lee, H. McKinzie, D. S. Tannhauser, and A. Wold, J of App. Phys. **40**, 602 (1969).
12. R. H. Friend, R. F. Frindt, A. J. Grant, A.D. Yoffe and D. Jerome, J. Phys. C: Solid State Phys. **10**, 1013 (1977).
13. T. E. Kidd, T. Miller, M. Y. Chou and T.C. Chiang, Phys. Rev. Lett. **88**, 226402 (2002).
14. K. C. Woo, F. C. Brown, W. L. McMillan, R. J. Miller, M. J. Schaffrman and M. P. Sears, Phys. Rev. B **14**, 3242 (1976).
15. E. Morosan, H. W. Zandbergen, B. S. Dennis, J. W. G. Bos, Y. Onose, T. Klimczuk, A. P. Ramirez, N. P. Ong, and R. J. Cava, Nature phys. **2**, 544 (2006).
16. E. Morosan, K. E. Wagner, Liang L. Zhao, Y. Hor, A. J. Williams, J. Tao, Y. Zhu, and R. J. Cava, Phys. Rev. B **81**, 094524 (2010).
17. D. J. Tozer, and M. J. Peach Physical Chemistry Chemical Physics, **16**(28), 14333 (2014).
18. Abhilasha Saini, Kapil Kumar, M. M. Sharma, R. P. Aloysius & V. P. S. Awana, J Supercond Nov Magn **35**, 1383 (2022).
19. Ranu Bhatt, Miral Patel, Shovit Bhattacharya, RanitaBasu, Sajid Ahmad, Pramod Bhatt, A K Chauhan, M Navneethan, Y Hayakawa, Ajay Singh, D K Aswal and S K Gupta, J. Phys. Cond. Mat. **26**, 445002 (2014).
20. D. Qian, D. Hsieh, L. Wray, E. Morosan, N. L. Wang, Y. Xia, R. J. Cava, and M. Z. Hasan, Phys. Rev. Lett. **98**, 117007 (2007).
21. P. Behera, SumitBera, M. M. Patidar, and V. Ganesan, AIP Conference Proceedings **2100**, 020113 (2019).
22. Lifei Sun, Chuanhui Chen, Dr. Qinghua Zhang, Christian Sohrt, Tianqi Zhao, Dr. Guanchen Xu ,Jinghui Wang, Prof. Dong Wang, Prof. Kai Rossnagel, Prof. Lin Gu, Prof. ChenggangTao, Prof. Liying Jiao, Angewandte Chemie **56**, 8981 (2019).
23. Matthias M. May, Christine Brabetz, Christoph Janowitz and Recardo Manzke, Journal of Electron Spectroscopy and Related Phenomena **184**, 180 (2011).
24. M. Calandra, and F. Mauri, Phys. Rev. Lett. **106**, 196406 (2011).
25. Q. Hu, J. Y. Liu, Q. Shi, F. J. Zhang, Y. Zhong, L. Lei, and R. Ang, EPL, **135**(5), 57003 (2021).
26. W. Liu, A. Luo, G. Zhong, J. Zou, and G. Xu, Phys. Rev. Research **4**, 023127 (2022).
27. P. Giannozzi, S. Baroni, N. Bonini, M. Calandra, R. Car, C. Cavazzoni, D. Ceresoli, G. L. Chiarotti, M. Cococcioni, I. Dabo, A. Dal Corso, S. De Gironcoli, S. Fabris, G. Fratesi, R. Gebauer, U. Gerstmann, C. Gougoussis, A. Kokalj, M. Lazzeri, L. Martin-Samos, N. Marzari, F. Mauri, R. Mazzarello, S. Paolini, A. Pasquarello, L. Paulatto, C. Sbraccia, S. Scandolo, G. Sclauzero, A. P. Seitsonen, A. Smogunov, P. Umari, and R. M. Wentzcovitch, J. Phys. Condens. Matter **21**, 395502 (2009).
28. P. Giannozzi, O. Andreussi, T. Brumme, O. Bunau, M. Buongiorno Nardelli, M. Calandra, R. Car, C. Cavazzoni, D. Ceresoli, M. Cococcioni, N. Colonna, I. Carnimeo, A. Dal Corso, S. De Gironcoli, P. Delugas, R. A. Distasio, A. Ferretti, A. Floris, G. Fratesi, G. Fugallo, R. Gebauer, U. Gerstmann, F. Giustino, T. Gorni, J. Jia, M. Kawamura, H. Y. Ko, A. Kokalj, E. Kücükbenli, M. Lazzeri, M. Marsili, N. Marzari, F. Mauri, N. L. Nguyen, H. V. Nguyen, A. Otero-De-La-Roza, L. Paulatto, S. Poncé, D. Rocca, R. Sabatini, B. Santra, M. Schlipf, A. P. Seitsonen, A. Smogunov, I. Timrov, T. Thonhauser, P. Umari, N. Vast, X. Wu, and S. Baroni, J. Phys. Condens. Matter **29**, 465901 (2017).
29. Y. Hinuma, G. Pizzi, Y. Kumagai, F. Oba, and I. Tanaka, Comput. Mater. Sci. **128**, 140 (2017).

Fig. 1(a)
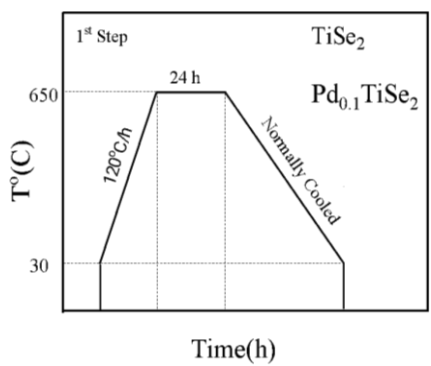

Fig. 1(b)
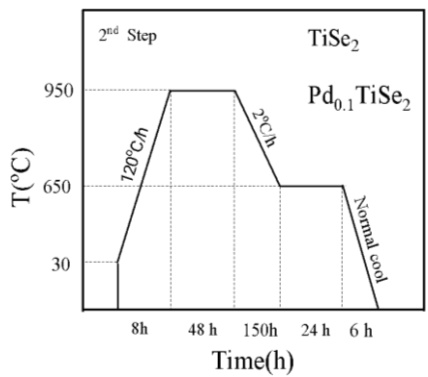

Fig. 2(a)
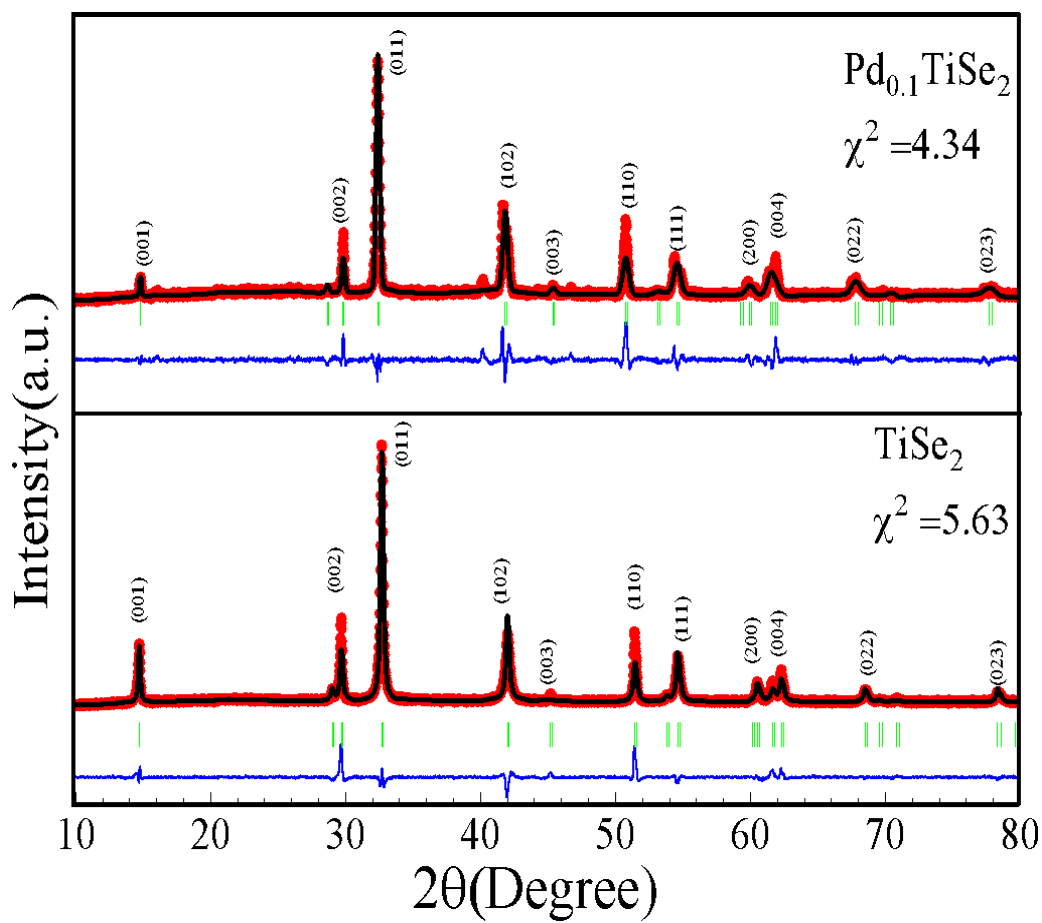

Fig.2 (b)

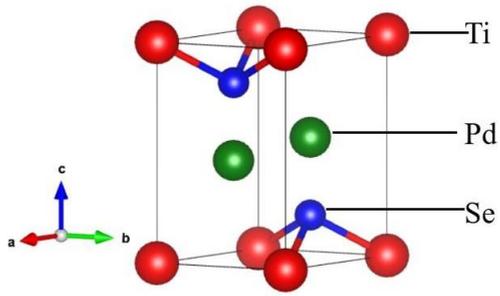

Fig. 2 (c)

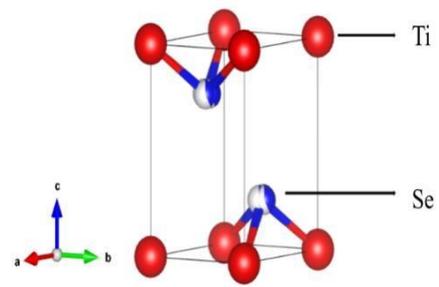

Fig.3 (a)

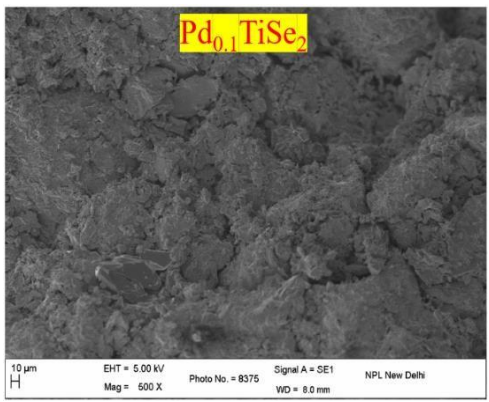

Fig. 3 (b)

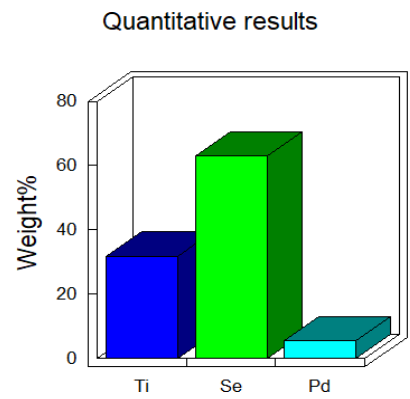

Fig.4

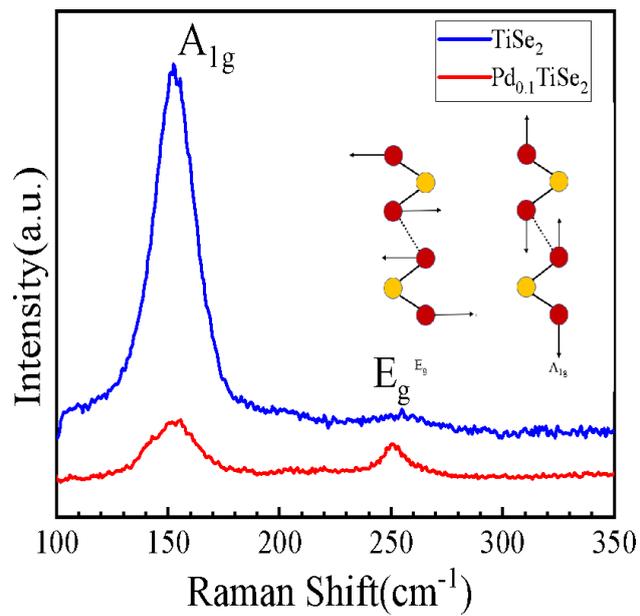

Fig.5 (a), 5(b) and 5(c)

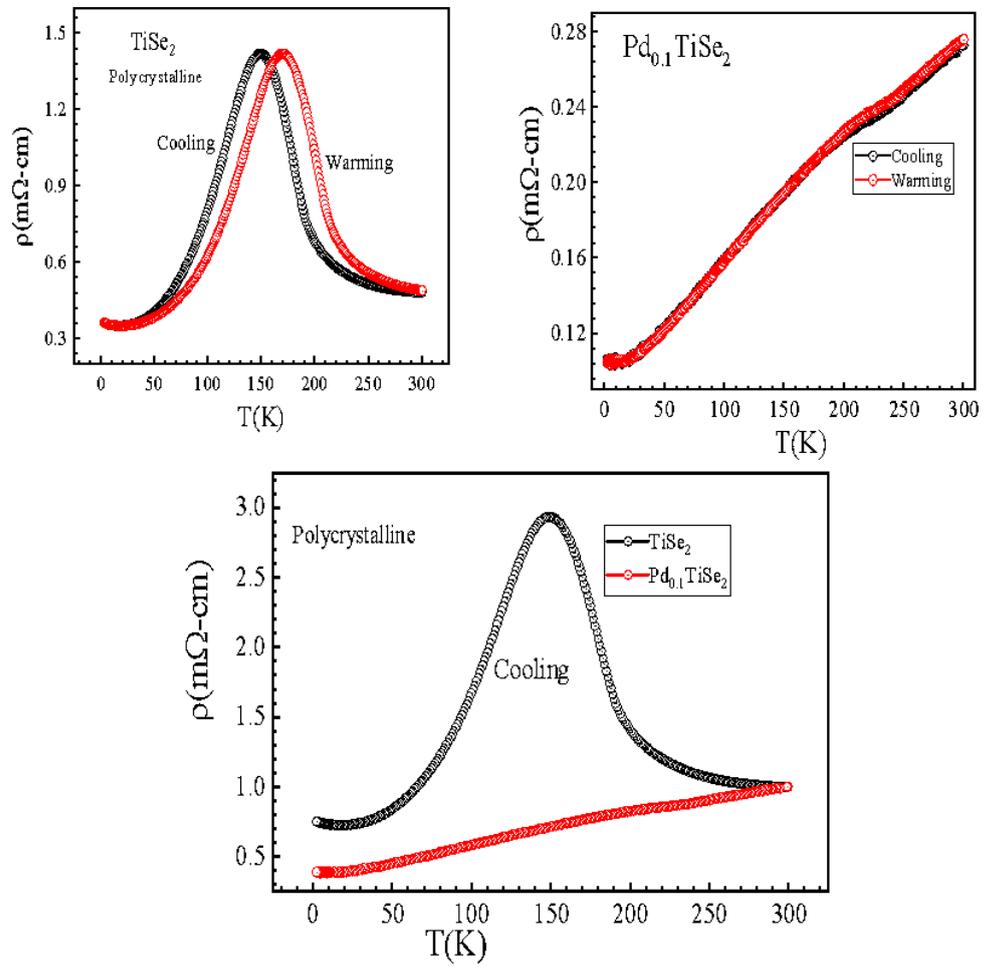

Fig. 6

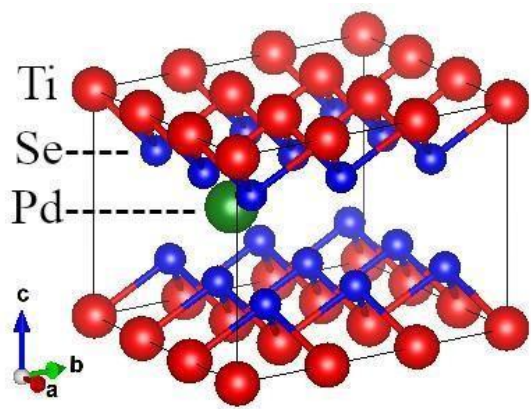

Fig. 7 (a)

Fig. 7(b)

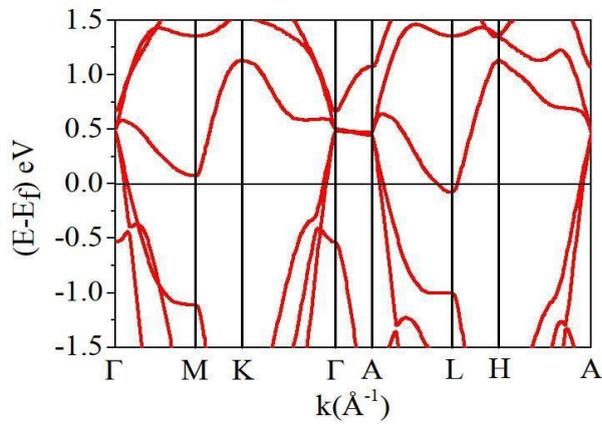
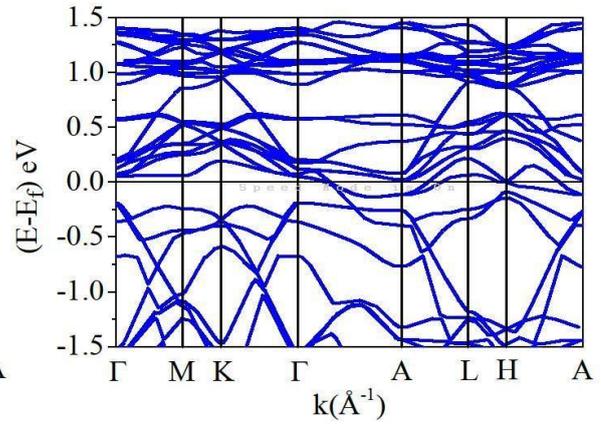

Fig. 7 (c)

Fig. 7(d)

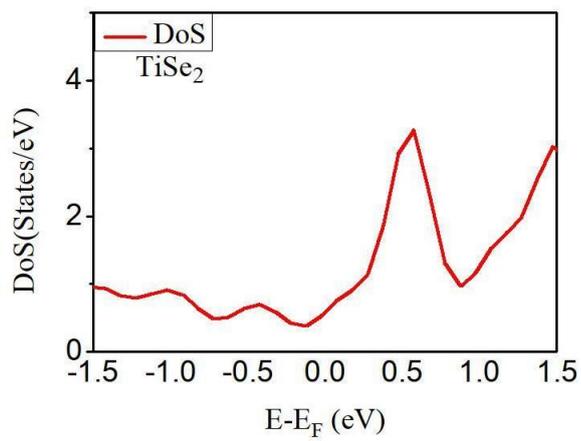
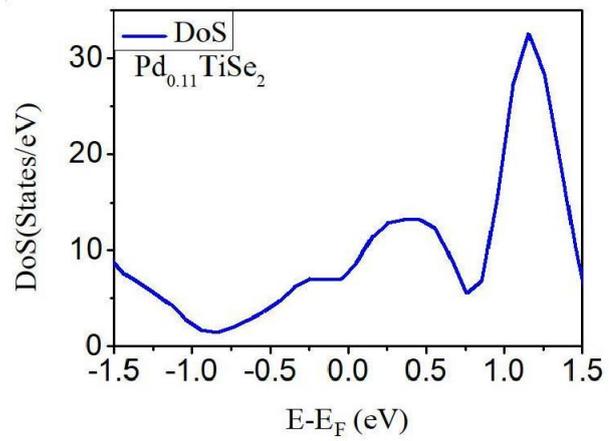